\documentclass{article}
\topmargin - 1.5 cm 
\begin{document}

\begin{center}
{\large{\bf The alternative model of the spherical oscillator}}
\end{center}

\begin{center}
{\bf Levon Mardoyan}
\end{center}
\begin{center}
Yerevan State University\\
Alex Manougian str., 1, 0025 Yerevan, Armenia\\
E-mail: mardoyan@ysu.am
\end{center}

\begin{abstract}
The quasiradial wave functions and energy spectra of the alternative
model of spherical oscillator on the $D$-dimensional sphere and
two-sheeted hyperboloid are found.
\end{abstract}


\section{Introduction}

The spherical oscillator was suggested by Higgs \cite{Higgs, Leemon}.
The $D$-dimensional spherical oscillator is defined by the potential
\begin{equation} \label{EQ1}
V_{S^D} = \frac{\omega^2}{2}\frac{x_\mu x_\mu}{x_0^2}, \qquad
\mu=1,2,\ldots,D,
\end{equation}
where $x_0,\,\,x_\mu$ are the Euclidean coordinates of the ambient
space ${\rm I \!R}^{D+1}$: $x_0^2+x_\mu x_\mu =r_0^2$ for D-dimensional
sphere and $x_0^2-x_\mu x_\mu =r_0^2$ for $D$-dimensional two-sheeted
hyperboloid. (We use a system of units in which the reduced mass $m$
and Planck constant $\hbar$ satisfy $m=\hbar=1$.)  The spherical
oscillator (\ref{EQ1}) on the $D$-dimensional sphere and two-sheeted
hyperboloid is considered in \cite{Kalnins} in detail.

The oscillator problem on spheres and pseudospheres was discussed from
many point of view in
\cite{Bonatos, Grocshe1, Grocshe2, Grocshe3,
Kalnins1, Kalnins2, Mardoyan1}.

The alternative model of spherical oscillator, which was suggested
in our previous papers \cite{Bellucci1, Mardoyan2}, is defined by
the potential
\begin{equation} \label{EQ2}
V_S^D = 2\omega^2r_0^2\frac{r_0 - x_0}{r_0+x_0}
\end{equation}
on the $D$-dimensional sphere, and
\begin{equation} \label{EQ3}
V_H^D = 2\omega^2r_0^2\frac{x_0-r_0}{x_0+r_0}
\end{equation}
on the $D$-dimensional two-sheeted hyperboloid.

The two-dimensional case of the oscillator potentials (\ref{EQ2}) and
(\ref{EQ3}) was considered in \cite{Bellucci2, Bellucci3}.

\section{Quasiradial function on $D$-sphere}

The Schr\"odinger equation describing the nonrelativistic quantum motion
in the $D$-di\-men\-sio\-nal curved space has the following form:
\begin{equation} \label{EQ4}
{\hat H}\Psi = \left[-\frac{1}{2}\Delta_{LB} +
V\left({\vec x}\right)\right]\Psi = E\Psi,
\end{equation}
where the Laplace-Beltrami operator in arbitrary curvilinear coordinates
$\xi_\mu$ is
$$
\Delta_{LB}=\frac{1}{\sqrt g}\frac{\partial}{\partial \xi_\mu}
\left(g^{\mu \nu}{\sqrt g}\frac{\partial}{\partial \xi_\mu}\right),
\qquad g=detg_{\mu \nu}, \qquad g_{\alpha \mu}g^{\mu \beta} =
\delta_\alpha^\beta.
$$
In the hyperspherical coordinates
\begin{eqnarray*}
x_0 &=& r_0\cos\chi, \\
x_1 &=& r_0\sin\chi \cos\theta_1, \\
x_2 &=& r_0\sin\chi \sin\theta_1 \cos\theta_2, \\
\vdots \\
x_{D-1} &=& r_0\sin\chi \sin\theta_1 \sin\theta_2 \cdots \sin\theta_{D-2}
\cos\varphi , \\
x_D &=& r_0\sin\chi \sin\theta_1 \sin\theta_2 \cdots \sin\theta_{D-2}
\sin\varphi ,
\end{eqnarray*}
where $\chi, \theta_1, \ldots, \theta_{D-2} \in [0,\pi]$,
$\varphi \in [0,2\pi)$, the oscillator potential (\ref{EQ2}) reads
\begin{equation} \label{EQ5}
V_S^D = 2\omega^2r_0^2\tan^2\frac{\chi}{2}.
\end{equation}
The Schr\"odinger equation (\ref{EQ4}) for the potential (\ref{EQ5})
may be solved by searching for a wave function in the form
$$
\Psi\left(\chi,\theta_1,\ldots,\theta_{D-2},\varphi\right)=
R(\chi)\,Y_{Ll_1l_2 \ldots l_{D-2}}
\left(\theta_1,\ldots,\theta_{D-2},\varphi\right),
$$
where $l_i$ are the angular hypermomenta and $L$ is total angular
momentum, and the hyperspherical function $Y_{Ll_1l_2 \ldots l_{D-2}}
\left(\theta_1,\ldots,\theta_{D-2},\varphi\right)$ is the solution of the
Laplace-Beltrami eigenvalue equation on the ($D$-1)-dimensional sphere.
After the separation of variables in (\ref{EQ4}) we obtain the quasiradial
equation
$$
\frac{1}{\left(\sin\chi\right)^{D-1}}\frac{\partial}{\partial \chi}
\left[\left(\sin\chi\right)^{D-1}\frac{\partial R}{\partial \chi}\right]
+\left[2r_0^2E - \frac{L(L+D-2)}{\sin^2\chi}-4\omega^2r_0^4
\tan^2\frac{\chi}{2}\right]R=0.
$$
Using the substitution
$$
R(\chi) = \left(\sin\chi\right)^{-\frac{D-1}{2}}Z(\chi)
$$
we find the P\"oschl-Teller type equation
\begin{equation} \label{EQ6}
\frac{d^2 Z}{d\xi^2} +\left[\epsilon -\frac{\nu^2-\frac{1}{4}}{\cos^2\xi}
-\frac{\left(L+\frac{D-2}{2}\right)^2-\frac{1}{4}}{\sin^2\xi}\right]Z=0,
\end{equation}
where $\xi=\frac{\chi}{2} \in \left[0, \frac{\pi}{2}\right]$, and
$$
\epsilon = 8r_0^2E+\left(D-1\right)^2+16\omega^2r_0^4, \qquad
\nu = \sqrt{\left(L+\frac{D-2}{2}\right)^2+16\omega^2r_0^4}.
$$
The solution of Eq. (\ref{EQ6}) regular for
$\xi \in \left[0, \frac{\pi}{2}\right]$ and expressed in terms
of the hypergeometric function is \cite{Flugge}
\begin{eqnarray}
R^{D}_{n_rL\nu}\left(\chi\right) = C^{D}_{n_rL\nu}
\left(\sin\frac{\chi}{2}\right)^L
\left(\cos\frac{\chi}{2}\right)^{\nu-\frac{D}{2}+1} \times \\ \nonumber
\label{EQ7} \\
\times {_2}F_1\left(-n_r,\,n_r+L+\nu+\frac{D}{2};\,L+\frac{D}{2};\,
\sin^2\frac{\chi}{2}\right), \nonumber
\end{eqnarray}
and the $\epsilon$ is quantized as
$$
\epsilon=\left(2n_r+L+\nu+\frac{D}{2}\right)^2,
$$
where $n_r=0,1,2,\ldots$ is a "quasiradial" quantum number.
The eigenvalues $E$ are given by
\begin{equation}\label{EQ8}
E_N^D=\frac{1}{8r_0^2}\left[(N+1)(N+D)+(2\nu-1)
\left(N+\frac{D}{2}\right)+L(L+D-2)-\frac{D}{2}(D-1)\right],
\end{equation}
where $N=2n_r+L=0,1,2,\ldots$ is the principal quantum number.

For the quasiradial wave function $R^{D}_{n_rL\nu}\left(\chi\right)$
we choose the normalization condition
$$
r_0^D\,\int\limits_{0}^{\pi}\,
\left|R^{D}_{n_rL\nu}\left(\chi\right)\right|^2\,
\left(\sin\chi\right)^{D-1}\,d\chi =1
$$
and find:
\begin{equation}\label{EQ9}
C^{D}_{n_rL\nu}=\sqrt{\frac{\left(2n_r+L+\nu+\frac{D}{2}\right)
\Gamma\left(n_r+L+\nu+\frac{D}{2}\right)
\Gamma\left(n_r+L+\frac{D}{2}\right)}
{2^{D-1}\,r_0^D\,\left(n_r\right)!\,
\Gamma\left(n_r+\nu+1\right)
\left[\Gamma\left(L+\frac{D}{2}\right)\right]^2}}.
\end{equation}

In the limit $r_0 \to \infty$, $\chi \to 0$ and $\chi r_0 \sim r$ - fixed
and $\nu \sim 4\omega r_0^2$, we see that
\begin{equation}\label{EQ10}
\lim\limits_{r_0 \to \infty}E_N^D = \omega \left(N+\frac{D}{2}\right)
\end{equation}
and
\begin{equation}\label{EQ11}
\lim\limits_{r_0 \to \infty}R_{NL\nu}^D(\chi) =
\frac{\omega^{\frac{L}{2}+\frac{D}{4}}}{\Gamma\left(L+\frac{D}{2}\right)}
\sqrt{\frac{2\Gamma\left(\frac{N+L+D}{2}\right)}
{\left(\frac{N-L}{2}\right)!}}
r^L\,e^{-\frac{\omega r^2}{2}}\,
F\left(-\frac{N-L}{2}; L+\frac{D}{2}; \omega r^2\right),
\end{equation}
where $F\left(a; c; x\right)$ is the confluent hypergeometric function.
Formula (\ref{EQ11}) coincides with the known formula for
$D$-dimensional flat radial wave functions \cite{MPSTA}.

\section{Oscillator on the $D$-dimensional hyperboloid}

The pseudospherical coordinates on the $D$-dimensional two-sheeted
hyperboloid: $x_0^2-x_1^2-x_2^2-x_D^2=r_0^2$, $x_0 \geq r_0$, are
\begin{eqnarray*}
x_0 &=& r_0\cosh\tau, \\
x_1 &=& r_0\sinh\tau \cos\theta_1, \\
x_2 &=& r_0\sinh\tau \sin\theta_1 \cos\theta_2, \\
\vdots \\
x_{D-1} &=& r_0\sinh\tau \sin\theta_1 \sin\theta_2 \cdots \sin\theta_{D-2}
\cos\varphi , \\
x_D &=& r_0\sinh\tau \sin\theta_1 \sin\theta_2 \cdots \sin\theta_{D-2}
\sin\varphi ,
\end{eqnarray*}
where $\tau \in [0,\infty)$. Variables in the Schr\"odinger equation
(\ref{EQ4}) may be separated for oscillator potential (\ref{EQ3})
which in the pseudospherical coordinates has the form
$$
V_H^D = 2\omega^2r_0^2\,\tanh^2\frac{\tau}{2},
$$
by the ansatz
$$
\Psi\left(\tau,\theta_1,\ldots,\theta_{D-2},\varphi\right)=
R(\tau)\,Y_{Ll_1l_2 \ldots l_{D-2}}
\left(\theta_1,\ldots,\theta_{D-2},\varphi\right),
$$
where, as in the previous case $l_i$, are the angular hypermomenta and $L$
is the total angular momentum, and the hyperspherical function
$Y_{Ll_1l_2 \ldots l_{D-2}}
\left(\theta_1,\ldots,\theta_{D-2},\varphi\right)$ is the solution of the
Laplace-Beltrami eigenvalue equation on the ($D$-1)-dimensional sphere.
After separation of variables in (\ref{EQ4}) we find the quasiradial
equation
$$
\frac{1}{\left(\sinh\tau\right)^{D-1}}\frac{\partial}{\partial \tau}
\left[\left(\sinh\tau\right)^{D-1}\frac{\partial R}{\partial \tau}\right]
+\left[2r_0^2E - \frac{L(L+D-2)}{\sinh^2\tau}-4\omega^2r_0^4
\tanh^2\frac{\tau}{2}\right]R=0.
$$
Using now the substitution
$$
R(\tau) = \left(\sinh\tau\right)^{-\frac{D-1}{2}}Z(\tau)
$$
we come to the equation
\begin{equation} \label{EQ12}
\frac{d^2 Z}{d\rho^2} +\left[\epsilon -\frac{\nu^2-\frac{1}{4}}{\cosh^2\rho}
-\frac{\left(L+\frac{D-2}{2}\right)^2-\frac{1}{4}}{\sinh^2\rho}\right]Z=0,
\end{equation}
where $\rho=\frac{\tau}{2} \in \left[0, \infty\right)$, and
$\epsilon = 8r_0^2-\left(D-1\right)^2-16\omega^2r_0^4.$

Thus, the oscillator problem on the two-sheeted hyperboloid is described
by the modified P\"oschl-Teller equation and, unlike the oscillator
equation on the sphere which has only a discrete spectrum,
equation (\ref{EQ12}) possesses both bound and unbound states.

The discrete quasiradial wave function regular on the line
$\tau \in [0, \infty)$ and normalized by the condition
$$
r_0^D\,\int\limits_{0}^{\infty}\,
\left|R^{D}_{n_rL\nu}\left(\tau\right)\right|^2\,
\left(\sinh\tau\right)^{D-1}\,d\tau =1
$$
has the form
\begin{eqnarray}
R^{D}_{n_rL\nu}\left(\tau\right) =
\frac{1}{\Gamma\left(L+\frac{D}{2}\right)}
\sqrt{\frac{\left(\nu-2n_r-L-\frac{D}{2}\right)
\Gamma\left(\nu-n_r\right)\Gamma\left(n_r+L+\frac{D}{2}\right)}
{2^{D-1}\,r_0^D\,\left(n_r\right)!\,
\Gamma\left(\nu-n_r-L-\frac{D}{2}+1\right)
}} \times \\ \nonumber
\label{EQ13} \\
\times \left(\sinh\frac{\tau}{2}\right)^L
\left(\cosh\frac{\tau}{2}\right)^{2n_r-\nu-\frac{D}{2}+1}
\times {_2}F_1\left(-n_r,\,-n_r+\nu;\,L+\frac{D}{2};\,
\tanh^2\frac{\tau}{2}\right), \nonumber
\end{eqnarray}
with the "quasiradial" quantum number
$n_r=0,1,2,\ldots,\left[\frac{1}{2}\left(\nu-L-\frac{D}{2}\right)\right]$.
The $\epsilon$ is quantized by
$$
\epsilon=-\left(2n_r+L-\nu+\frac{D}{2}\right)^2,
$$
and the energy spectrum for the alternative model of quantum
spherical oscillator on the $D$-dimensional two-sheeted
hyperboloid takes the value
\begin{equation}\label{EQ14}
E_N^D=\frac{1}{8r_0^2}\left[(2\nu-1)\left(N+\frac{D}{2}\right)
-N(N+D-1)-L(L+D-2)+\frac{D}{2}(D-1)\right].
\end{equation}
Here $N=2n_r+L$ is the principal quantum number and the bound
state solution is possible only for
$$
0 \leq N \leq \left[\nu-\frac{D}{2}\right].
$$
In the contractio limit $r_0 \to \infty$, $\tau \sim r/r_0$
and $\nu \sim 4\omega r_0^2$, we see that the continuous spectrum
vanishes while the discrete spectrum is infinite, and it is
easy to reproduce the oscillator energy spectrum (\ref{EQ10}) and
wave function (\ref{EQ11}).

\section*{Acknowledgments.}
The author is grateful to Dr. A.~Nersessian for many dicussions and to the
Organizing Committee and personally to Prof. C.~Burdik for the invitation
to the XVI-th International Colloquium on {\em Integrable Systems and
Quantum Symmetries} and for warm hospitality in Prague.
This work was partially supported by the grant NFSAT-CRDF ARP1-3228-YE-04.

\end{document}